\documentclass[10pt,letterpaper,twocolumn]{article} 

\usepackage{ol2}
\usepackage[draft]{hyperref}
\usepackage{amsmath}

\begin{document}

\twocolumn[ 

\title{Monolithic whispering-gallery mode resonators with vertically coupled integrated bus waveguides}

\author{Mher Ghulinyan,$^{1,*}$ Romain Guider,$^2$ Georg Pucker,$^1$ and Lorenzo Pavesi$^{2}$}

\address{
$^1$Advanced Photonics \& Photovoltaics, Fondazione Bruno Kessler, via Sommarive 18, Povo (TN), Italy\\
$^2$Nanoscience Lab., Department of Physics, University of Trento, via Sommarive 14, Povo (TN), Italy\\
$^*$Corresponding author: ghulinyan@fbk.eu
}

\begin{abstract}
We report on the realization and optical characterization of a CMOS-compatible silicon-based microresonator/waveguide coupled system, fully
integrated on a silicon chip. The device uses a vertical coupling scheme between the resonator and a buried strip waveguide. We demonstrate that
its high optical quality follows from the accurate planarization of the waveguide topography.
More importantly, we demonstrate a wafer-scale mass
fabrication of freestanding planar resonators suspended in air and coupled to the integrated bus waveguides. A nanometer control of the coupling distances allows for a precise and selective excitation of different mode families of the resonator. This opens the door for the realization of stable all-integrated complex resonator systems for optomechanical and metrological applications, with the potential to substitute the nowadays intensive use of complicated fiber-taper coupling schemes.
\end{abstract}


] 

\noindent One of the most important requirements for planar Whispering-gallery mode (WGM) resonators is a high refractive index contrast between
the cavity and the surrounding media. For this reason, a quasi-freestanding resonator embedded in air is the preferred solution \cite{cuttedge}.
In practice, this is achieved by underetching the resonator base to form a small diameter pedestal. Most of current cutting-edge experiments
with this kind of optically passive and freestanding resonators regularly uses tapered fibers to probe the system
\cite{optomech1,optomech2,phonlaser}. In this scheme, a tapered fiber is laterally approached to the cavity to a distance that allows an evanescent field coupling. Positioning the tapered fiber requires the use of piezoelectric controllers to provide nanometric coupling gap. This,
however, results in unstable experimental conditions. The lack of a freestanding WGM cavity system monolithically integrated with an
input/output waveguide reduces the chances for the applications of these devices \cite{raised}.

In planar lightwave circuits, the \emph{vertical} coupling of a WGM resonator to the bus waveguide (the resonator and the bus waveguide lay on different
planes) has several advantages with respect to the \emph{lateral} coupling (the resonator and the bus waveguide lay on the same plane). In fact,
vertical coupling:
\begin{enumerate}
\item[(i)] enables denser device integration; \item[(ii)] avoids the use of ultra-high resolution lithography, such as deep-UV or electron beam lithography
to control the gap width at the nanometer level; \item[(iii)] permits to use different materials and thicknesses for the resonator and the waveguide;
\item[(iv)] allows the selective exposure of the resonator or the bus waveguide to the ambient for sensing applications.
\end{enumerate}
These advantages are counterbalanced by delicate fabrication processes. The most delicate process step is the planarization of the strip
waveguide cladding which precedes the microdisk fabrication. Different solutions to solve partially this issue have been reported in the literature. For example, in
\cite{i3e1999} channel waveguides have been buried in SiO$_2$ by using a chrome lift-off technique, while in \cite{nonlin-hydex,opo-hydex} a simultaneous deposition/removal procedure of a deuterium rich oxynitride (SiO$_x$N$_y$:D) cladding was
used. CMOS-compatible heterogenous III-V-on-SOI (Silicon on Insulator) technology uses molecular or adhesive die-to-wafer bonding to couple
vertically III-V microring lasers to silicon waveguides \cite{ghent1,ghent2,leti}. As an adhesive a polymer, which undergoes a pressure-assisted
curing, is used with a rough control of the coupling.

Here we report on the realization and optical testing of a CMOS-compatible vertically coupled system using standard silicon microfabrication
technology. We planarized a silica cladding over the strip waveguides and simultaneously defined the vertical coupling gap.
The coupling layer thickness was accurately controlled and a high degree of planarization (DOP) was achieved. Furthermore, we have developed a technology to realize freestanding WGM resonators suspended over the bus
waveguides, thus, demonstrating an effective air coupling in an integrated system. Waveguide transmission measurements revealed quality factors
of several 10,000's from very first devices.

\begin{figure}[t!]
\centerline{\includegraphics[width=7.2cm]{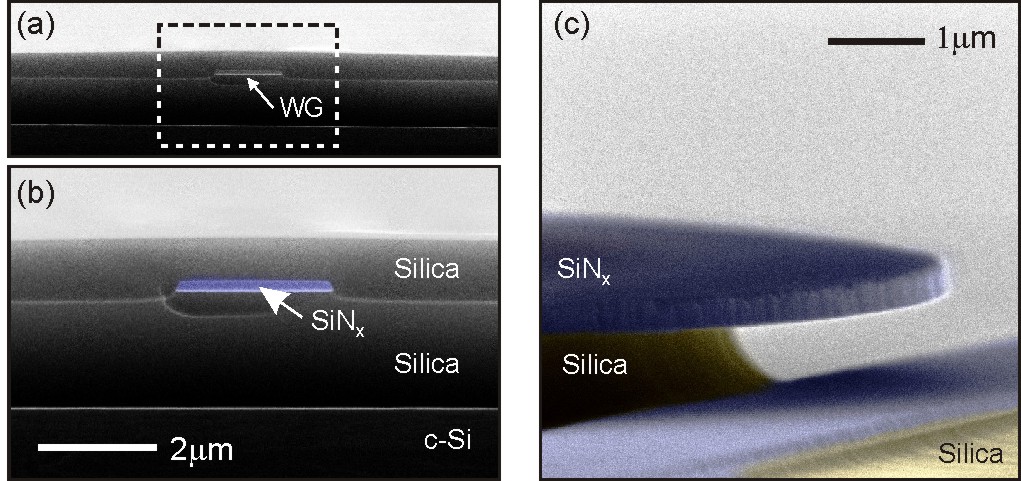}}
  \caption{
  SEM images (false-colored) of the device. (a) The cross-sectional image shows a quasi-full planarization of the  strip waveguide [DOP=90\%].
  (b) A close-up of the waveguide facet. (c) A tilted angle view of the vertical coupling zone between the waveguide and air suspended microdisk resonator. }
\label{sem}
  \end{figure}

\begin{figure}[t!]
\centerline{\includegraphics[width=7.5cm]{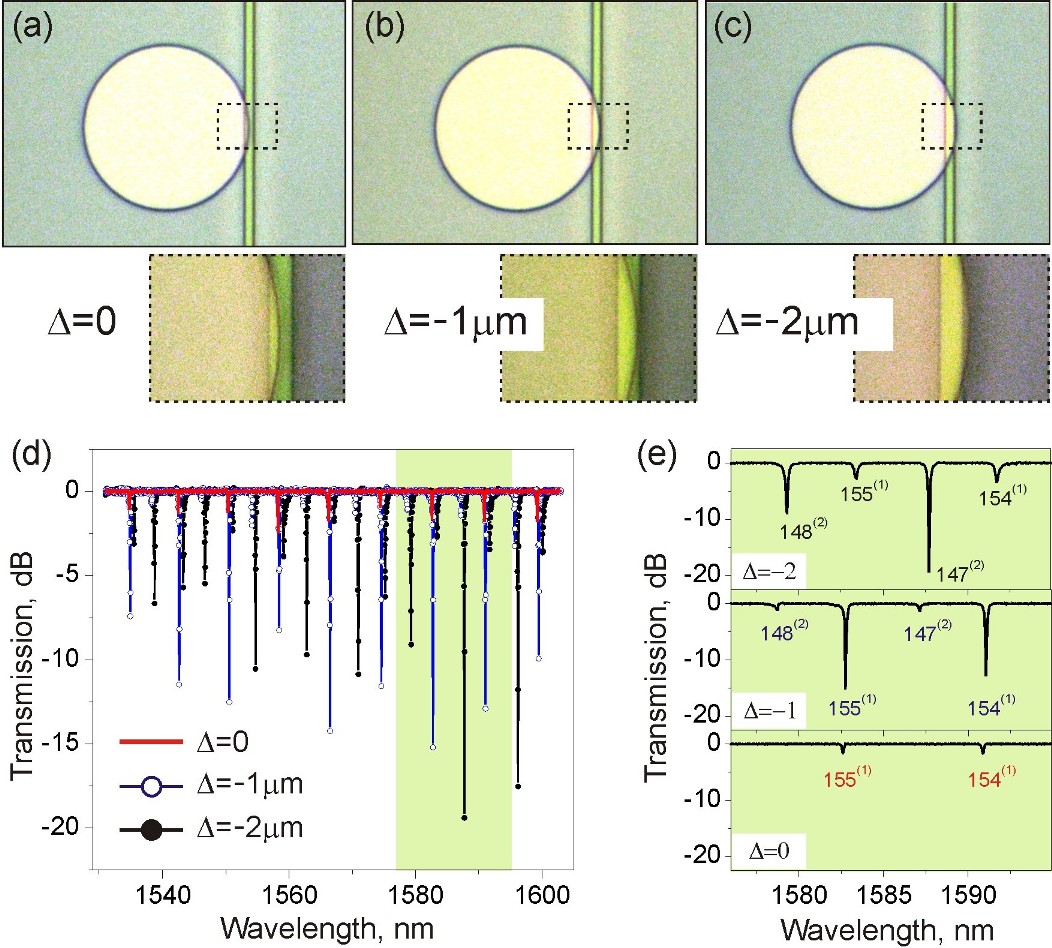}}
\caption{Top-view optical images of WGM microdisks vertically coupled to bus waveguides. The parameter $\Delta$ is the lateral
distance between the microdisk edge and the waveguide axis: (a) $\Delta$=0, (b) $-1\mu$m and (c) $-2\mu$m. (d) Normalized waveguide transmission
spectra over a wide wavelength range. (e) Zoom of the spectra around $\lambda\approx1585$~nm. The different resonances are labeled with their corresponding radial and azimuthal mode numbers.}
  \label{spectra}
  \end{figure}

The device fabrication process starts with a plasma enhanced chemical vapor deposition (PECVD) of a 2 $\mu$m-thick SiO$_2$ cladding layer on a crystalline Si wafer. Then, a 260 nm PECVD SiN$_x$ layer was deposited on top of the cladding, patterned lithographically and successively dry-etched to form strip waveguides. The dry etching procedure, accounting the cladding overetch, results in a step-height of $\sim$350~nm at the edges of the waveguide. Without planarization, such a topography will be transferred to the coupling silica spacer and, finally, to the resonator layer, resulting in an extremely lossy resonator. Therefore, we cladded the strip waveguides and planarized them with a silica layer. We adjusted the silica layer thickness using a wet etch in order to define the vertical coupling gap. Finally, a 350 nm thick PECVD SiN$_x$ layer was deposited, and WGM resonators were defined through standard photolithographic and dry etching steps. As a result of this process, monolithic microresonators vertically coupled to
buried waveguides were realized [Fig.~\ref{sem}(c) and Fig.~\ref{spectra}(a)--\ref{spectra}(c)].

The devices were characterized by measuring the bus waveguide transmission with a near-infrared tunable laser butt-coupled to the bus waveguide.
The signal polarization was controlled at the input and analyzed at the output. A beam splitter was used to monitor the transmitted mode in an infrared  CCD camera and to measure the signal intensity in a photodiode simultaneously.

Here we report on the results obtained from 50$\mu$m-diameter microdisk resonators. To demonstrate the potential of the vertical coupling method
we studied three WGM devices with the same coupling gap of 850 nm but different lateral alignments between the microdisk
and the bus waveguide [Fig.~\ref{spectra}(a)--\ref{spectra}(c)]. The alignment is described by the parameter $\Delta$ which measures the position of the disk
edge relative to the waveguide axis. Figure~\ref{spectra}(d) shows the transmission spectra from these devices \cite{note1}. A series of transmission dips
(cavity resonances) due to the coupling of the signal into the microdisk is observed. The homogenous distribution of cavity resonances
throughout the whole spectrum with a mean quality factor of $\bar{Q}\approx18,000$ confirms the good optical quality of the microdisks, i.e. of
the planarization process. Moreover, statistical analysis of transmission performed over many 50 $\mu$m resonators (same $\Delta$) from
different wafers resulted in a less than 1~nm spectral variation of WGM peak positions. Finite element analysis suggests that this corresponds
to a diameter variation of $\Delta d\approx40$~nm, which implies an excellent wafer scale mass fabrication of WGM resonators with a size
deviation of 0.08\% only.

In Fig.~\ref{spectra}(e) we show the high-resolution transmission spectra around $\lambda$=1585~nm for different $\Delta$ values. Different
spectra are observed even though the same coupling gap is used. Firstly, we performed finite element simulations in order to identify the
observed modal structure of microdisk (radial families, $p$, and azimuthal mode numbers $M$, labeled as $M^{(p)}$ hereafter). From the bottom
panel of Fig.~\ref{spectra}(e) we see that when $\Delta=0$, only the first radial mode family, with a free spectral range of $\approx8$~nm, is
excited through the waveguide ($154^{(1)}$,$155^{(1)}$). When $\Delta=-1~\mu$m [central panel of Fig.~\ref{spectra}(e)], this family is more
efficiently excited while modes with $p$=2, ($147^{(2)}$,$148^{(2)}$), appear in the spectrum. A larger overlap between the bus waveguide and the microdisk inverts the
relative intensities of the mode families [top panel, Fig.~\ref{spectra}(e)]. For this $\Delta$ value, $M^{(2)}$ modes are spatially aligned to the waveguide. This result shows a
further advantage of the vertical coupling scheme: it allows for selective excitation of different mode families which is impossible with
lateral coupling.

\begin{figure}[t!]
\centerline{\includegraphics[width=7.2cm]{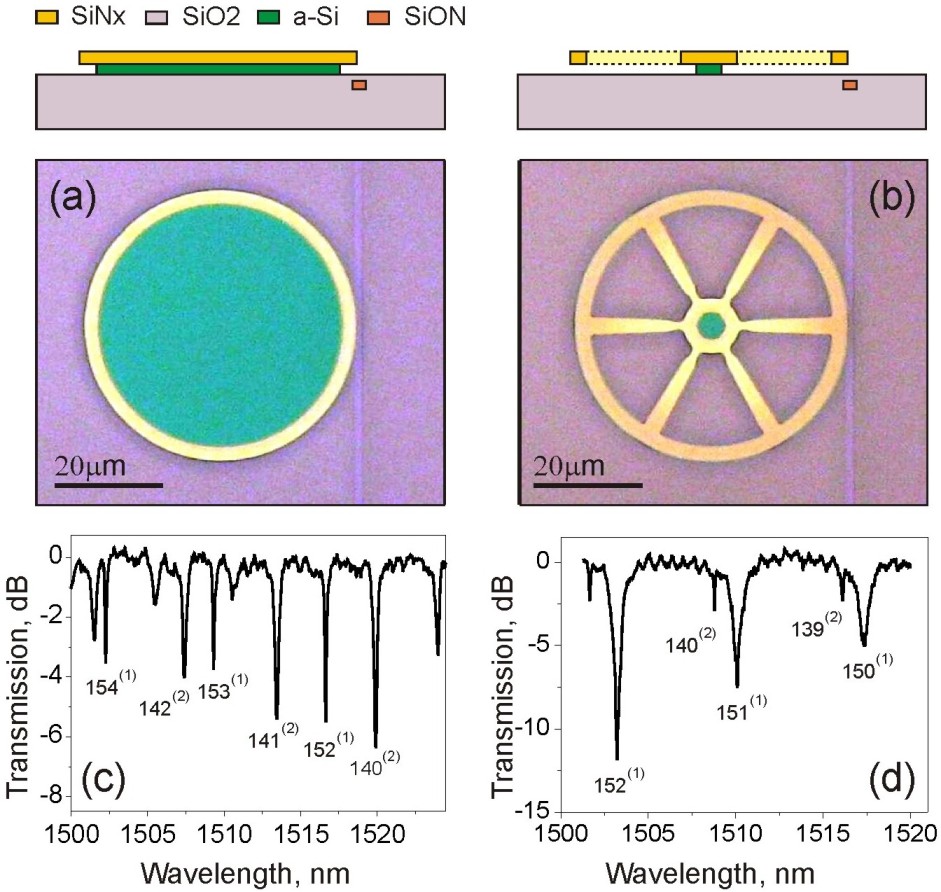}}
\caption{Free-standing WGM resonators with monolithically integrated bus waveguides. Optical micrographs of (a) a
raised microdisk resonator and (b) a spiderweb ring resonator with corresponding cross-sectional sketches (top panels). The transmission
spectra, plotted in (c) and (d), show the presence of WGM resonances with quality factors of the order of $(1\div2.5)\times10^4$ depending on
the coupling strength of various modes $M^{(p)}$.}
  \label{polyNit}
  \end{figure}

More importantly, the proposed approach enables the realization of freestanding WGM resonators suspended in air over the bus waveguides. This, for example, can
be achieved by a partial removal of silica under the resonator using a selective wet etch [Fig.~\ref{sem}(c)]. In specific cases, when a deep
underetch is required, one might free partially or completely the bus waveguide. If this is undesirable, the deposition of a sacrificial layer
between the silica and the resonator layers  can allow to preserve the bus waveguide cladding [see the cross-sectional sketches in Fig.~\ref{polyNit}]. In fact, the sacrificial layer can be removed isotropically through either a chemical wet or dry etching, leaving the cladding silica layer unaffected.

In Fig.~\ref{polyNit}(a) and \ref{polyNit}(b) examples of a freestanding WGM microdisk and a spiderweb ring resonator air-coupled to the bus waveguides are
shown. In this case, after the planarization, the silica layer was reduced to $\approx300$~nm and covered by a 250~nm-thick amorphous silicon
(a-Si) layer. After the deposition and definition of SiN$_x$ microresonators, the a-Si layer was selectively removed by using an isotropic
dry Si deep reactive ion etch. The underlaying silica layer remained unaffected. Thus, WGM resonators suspended in air over the cladded bus
waveguides were successfully realized. The transmission spectra of the suspended microresonators are plotted in Figs.~\ref{polyNit}(c) and \ref{polyNit}(d).
The resonant nature of the optical devices is clearly seen, and the various $M^{(p)}$ modes can be identified. We note that the vertical coupling gaps
for these first devices were not optimized. While the microdisk spectrum is characterized by narrow resonances, the spiderweb shows rather broad
first radial family modes. On one side, the spiderweb resonator possesses a lower effective index and therefore a weaker mode confinement than
the microdisk. On the other side, the residual film stress after the underetch bends the spiderweb towards the waveguide pushing it into a
stronger coupling regime. Both effects, therefore, result in lower quality factors for the spiderweb resonator, which could be avoided through
careful optimization.

In conclusion, we have demonstrated wafer-scale integration of a monolithic, CMOS-compatible microresonator/waveguide vertically coupled system
on a silicon chip. An easily accessible technological approach has been implemented for the realization of freestanding WGM resonators suspended
over the integrated bus waveguides with quality factors of $2\times10^4$ from very first devices. These solutions can be exploited not
only for the developments of compact Si-based lightwave circuits, but more importantly, can bring to the industrial market all-integrated
optomechanics with WGM resonators.

The authors wish to thank A. Pitanti and P. Bettotti for stimulating discussions and technical help. This work has been supported by NAoMI-FUPAT.


\begin{thebibliography}{99}

\bibitem{cuttedge}
A. B. Matsko, \emph{Practical Applications of Microresonators in Optics and Photonics}, (CRC Press, Taylor \& Francis Group, USA, 2009).

\bibitem{optomech2}
P. Del'Haye, A. Schliesser, O. Arcizet, T. Wilken, R. Holzwarth and T. J. Kippenberg, ``Optical frequency comb generation from a monolithic
microresonator," Nature \textbf{450,} 1214--1217 (2007).

\bibitem{optomech1}
T.J. Kippenberg and K.J. Vahala, ``Cavity Optomechanics: Back-Action at the Mesoscale," Science \textbf{321,} 1172--1176 (2008).

\bibitem{phonlaser}
I.S. Grudinin, H. Lee, O. Painter and K.J. Vahala, ``Phonon laser action in a tunable two-level system," Phys. Rev. Lett. \textbf{104,} 083901 (2010).

\bibitem{raised}
B. Redding, E. Marchena, T. Creazzo, S. Shi, and D.W. Prather, ``Comparison of raised-microdisk whispering gallery-mode characterization techniques," Opt. Lett. \textbf{35,} 998--1000 (2010).

\bibitem{i3e1999}
B.E. Little, S.T. Chu, W. Pan, D. Ripin, T. Kaneko, Y. Kokubun, and E. Ippen, ``Vertically Coupled Glass Microring Resonator Channel Dropping Filters," IEEE Photon. Tech. Lett. \textbf{11,} 215--217 (1999).

\bibitem{nonlin-hydex}
M. Ferrera, L. Razzari, D. Duchesne, R. Morandotti, Z. Yang, M. Liscidini, J.E. Sipe, S. Chu, B.E. Little and D.J. Moss, ``Low-power continuous-wave nonlinear optics in doped silica glass integrated waveguide structures," Nature Photon. \textbf{2,} 737--740 (2008).

\bibitem{opo-hydex}
L. Razzari, D. Duchesne, M. Ferrera, R. Morandotti, S. Chu, B.E. Little and D.J. Moss, ``CMOS-compatible integrated optical hyper-parametric oscillator," Nature Photon. \textbf{4,} 41--45 (2009).

\bibitem{ghent1}
G. Roelkens, D. Van Thourhout, R. Baets, R. N\"{o}tzel, and M. Smit, ``Laser emission and photodetection in an InP/InGaAsP layer integrated on and coupled to a Silicon-on-Insulator waveguide circuit," Opt. Express \textbf{14,} 8154--8159 (2006).

\bibitem{ghent2}
L. Liu, R. Kumar, K. Huybrechts, T. Spuesens, G. Roelkens, E.-J. Geluk, T. de Vries, P. Regreny, D. Van Thourhout, R. Baets, and G. Morthier, ``An ultra-small, low-power, all-optical flip-flop memory on a silicon chip," Nature Photon. \textbf{4,} 182--187 (2010).

\bibitem{leti}
P. Viktorovitch, B. Ben Bakir, S. Boutami, J. L. Leclercq, X. Letartre, P. Rojo-Romeo, C. Seassal, M. Zussy, L. Di Cioccio, and J. M. Fedeli, ``3D harnessing of light with 2.5D photonic crystals," Laser Photon. Rev. \textbf{4,} 401--413 (2010).

\bibitem{note1}
The raw spectra were normalized to the transmission spectra of a bare waveguide on the same chip.

\end{thebibliography}
\end{document}